\documentclass[
aip,
 amsmath,amssymb,
reprint,%
]{revtex4-1}
\usepackage{booktabs}
\usepackage{hhline}

\usepackage{siunitx}
\DeclareSIUnit\angstrom{\text{\AA}}

\usepackage{svg}
\usepackage{graphicx}% Include figure files
\usepackage{dcolumn}% Align table columns on decimal point
\usepackage{bm}% bold math
\usepackage{hyperref}
\hypersetup{
    unicode=false,          % non-Latin characters in Acrobat’s bookmarks
    pdftoolbar=true,        % show Acrobat’s toolbar?
    pdfmenubar=true,        % show Acrobat’s menu?
    pdffitwindow=false,     % window fit to page when opened
    pdfstartview={FitH},    % fits the width of the page to the window
    colorlinks=true,        % false: boxed links; true: colored links
    linkcolor=blue,         % color of internal links
    citecolor=blue,         % color of links to bibliography
    urlcolor=blue           % color of URL in references
}

\usepackage[utf8]{inputenc}
\usepackage[T1]{fontenc}
\usepackage{mathptmx}
\usepackage{etoolbox}

\usepackage{float}

\pdfoutput=1

\makeatletter
\def\@email#1#2{%
 \endgroup
 \patchcmd{\titleblock@produce}
  {\frontmatter@RRAPformat}
  {\frontmatter@RRAPformat{\produce@RRAP{*#1\href{mailto:#2}{#2}}}\frontmatter@RRAPformat}
  {}{}
}%
\makeatother
\begin{document}

\preprint{AIP/123-QED}

\title{Automatic Differentiation for Orbital-Free Density Functional Theory}

\author{Chuin Wei Tan}
\affiliation{Department of Materials Science and Metallurgy, University of Cambridge, 27 Charles Babbage Road, Cambridge CB3 0FS, United Kingdom}

\author{Chris J. Pickard}
\affiliation{Department of Materials Science and Metallurgy, University of Cambridge, 27 Charles Babbage Road, Cambridge CB3 0FS, United Kingdom}
\affiliation{Advanced Institute for Materials Research, Tohoku University, 2-1-1 Katahira, Aoba, Sendai 980-8577, Japan}

\author{William C. Witt*}
\email{wcw28@cam.ac.uk}
\affiliation{Department of Materials Science and Metallurgy, University of Cambridge, 27 Charles Babbage Road, Cambridge CB3 0FS, United Kingdom}

\date{\today}

\begin{abstract}
Differentiable programming has facilitated numerous methodological advances in scientific computing. Physics engines supporting automatic differentiation have simpler code, accelerating the development process and reducing the maintenance burden. Furthermore, fully-differentiable simulation tools enable direct evaluation of challenging derivatives---including those directly related to properties measurable by experiment---that are conventionally computed with finite difference methods. Here, we investigate automatic differentiation in the context of orbital-free density functional theory (OFDFT) simulations of materials, introducing PROFESS-AD. Its automatic evaluation of properties derived from first derivatives, including functional potentials, forces, and stresses, facilitates the development and testing of new density functionals, while its direct evaluation of properties requiring higher-order derivatives, such as bulk moduli, elastic constants, and force constants, offers more concise implementations compared to conventional finite difference methods. For these reasons, PROFESS-AD serves as an excellent prototyping tool and provides new opportunities for OFDFT.

\end{abstract}

\maketitle

\section{\label{intro} Introduction}

Automatic differentiation (AD) is a family of techniques for algorithmic computation of the derivatives of a function specified by a computer program. Its utility for the physical sciences was recognized soon after its introduction,\cite{first_AD,early_AD4DE} and it has remained a topic of interest for scientific computing in the decades since.\cite{ADIFOR, AUTO_DERIV, DNAD} However, its adoption has accelerated recently in conjunction with trends in machine learning and the addition of AD frameworks to more programming platforms.\cite{tensorflow, julia, pytorch, jax} In the paradigm of differentiable physics,\cite{differentiable_physics} AD aids the modelling of physical systems as varied as fluid dynamics\cite{autograd_fluids} and molecular dynamics.\cite{autograd_md_1,autograd_md_2} In the realm of electronic structure, AD has been used for Hartree-Fock basis set optimization\cite{qc_ad_2} and the computation of molecular properties depending on higher-order derivatives of Hartree-Fock or density functional theory (DFT) energies.\cite{dqc} It has also been used to compute arbitrary-order exchange-correlation functional derivatives to facilitate time-dependent DFT simulations.\cite{qc_ad_1} Further examples include DFT and quantum chemistry codes with AD capabilities\cite{kasim_xc, dick_fs_xc, DFTKjcon, differentiable_pyscf} and the "Kohn-Sham regularizer",\cite{ks_regularizer, ks_regularizer2} which improves machine-learned exchange-correlation functionals. \\

Orbital-free DFT (OFDFT)\cite{ofdft_rev_LC, ofdft_rev_KCT, ofdft_rev_WW, OFDFT_rev} is a promising sub-field that would benefit from a fully differentiable implementation. In contrast with conventional Kohn-Sham DFT (KSDFT),\cite{KS} whose computational cost increases cubically with system size for large systems, OFDFT offers \mbox{(quasi-)linear} scaling, allowing for million-atom simulations,\cite{pme3, ofdft-million-1, ofdft-million-2, DFTPy} as well as much faster simulations of moderate size. The reduced computational complexity is achieved because OFDFT determines the system energy directly from the electron density, bypassing any need for individual orbitals and thereby eliminating costs associated with orbital manipulation. Several relatively mature OFDFT codes are in use based on these principles, including PROFESS,\cite{profess1,profess2,profess3} ATLAS,\cite{ATLAS} DFTPy,\cite{DFTPy} and CONUNDrum.\cite{conundrum} However, while OFDFT is an exact theory, in practice it is typically less accurate than conventional KSDFT, suffering from limitations of local pseudopotentials and the noninteracting kinetic energy functional approximations necessary for the orbital-free framework. Overcoming these limitations remains an active area of research. \\

This manuscript demonstrates how differentiable programming can simplify and enhance OFDFT simulations. We begin in Section~\ref{background} with  background on the relevant AD strategies, as well as the general OFDFT formulation. In Section~\ref{PROFESS-AD_intro}, we introduce PROFESS-AD, an auto-differentiable OFDFT code. Finally, in Section~\ref{PROFESS-AD_applications}, we present applications demonstrating its utility for OFDFT users and developers.

\section{\label{background} Background and Theory}

\subsection{\label{autodifferentiation_section} Automatic Differentiation}

AD is based on the principle that all numerical computations are inherently compositions of a finite set of elementary operations, each of which has a known derivative.\cite{autodiff1} By tracking the elementary parts of a complicated calculation, it becomes possible to efficiently, in an automated fashion, combine elementary derivatives into more challenging derivatives using the chain rule. AD is exact and its speed is comparable to that of manually coded derivatives.\cite{autodiff2} In fact, the cost of evaluating a scalar-valued function with backward mode AD is bounded by a small multiple of the cost of the function evaluation itself.\cite{autodiff_cost1, autodiff_cost2}

\subsubsection{Backward Mode Automatic Differentiation}

\begin{figure}[htb]
\centering
\includegraphics[width=\linewidth]{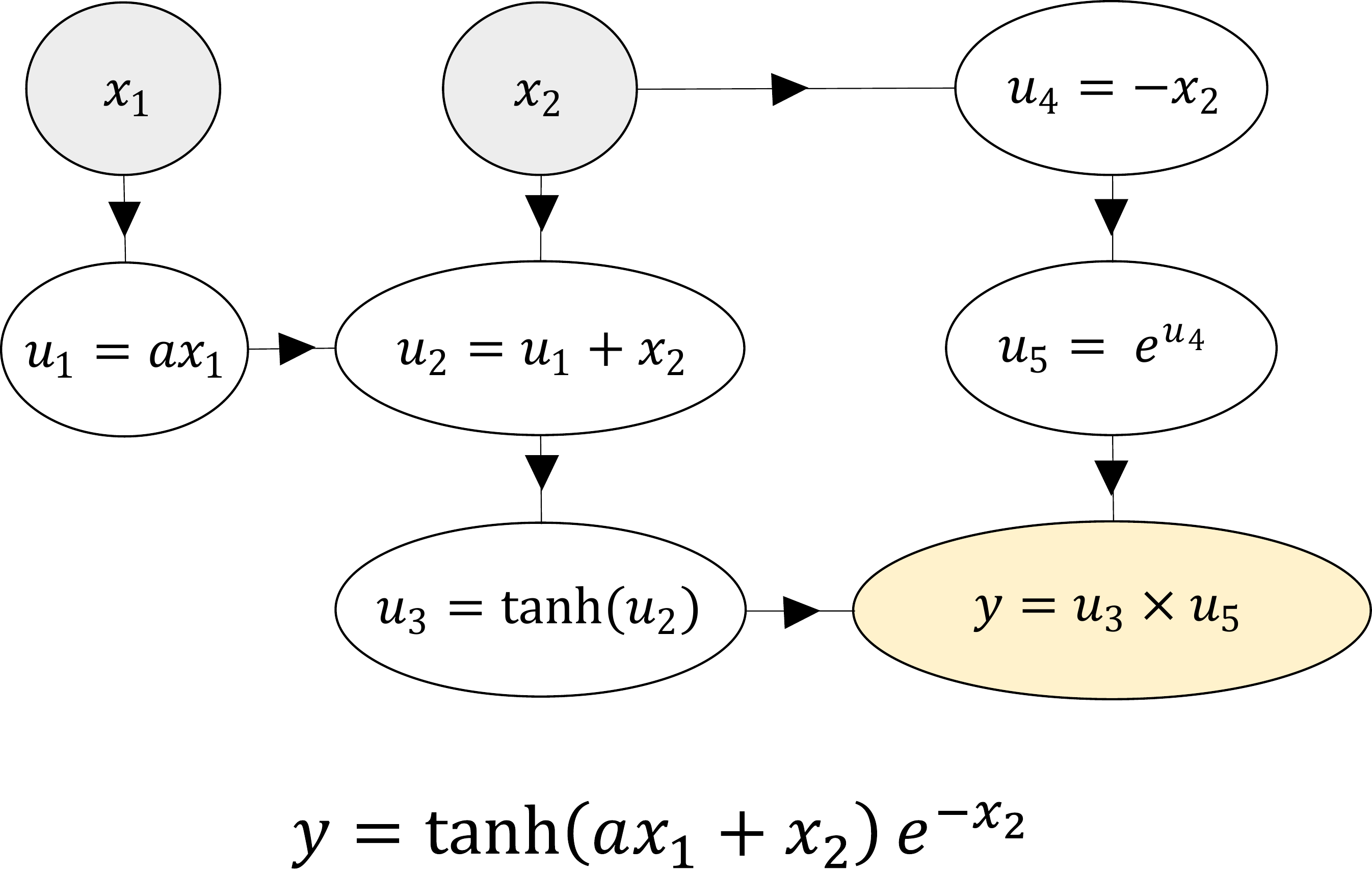}
\caption{Construction of a graph of elementary operations.}
\label{forward}
\end{figure}

\begin{figure}[htb]
\centering
\includegraphics[width=\linewidth]{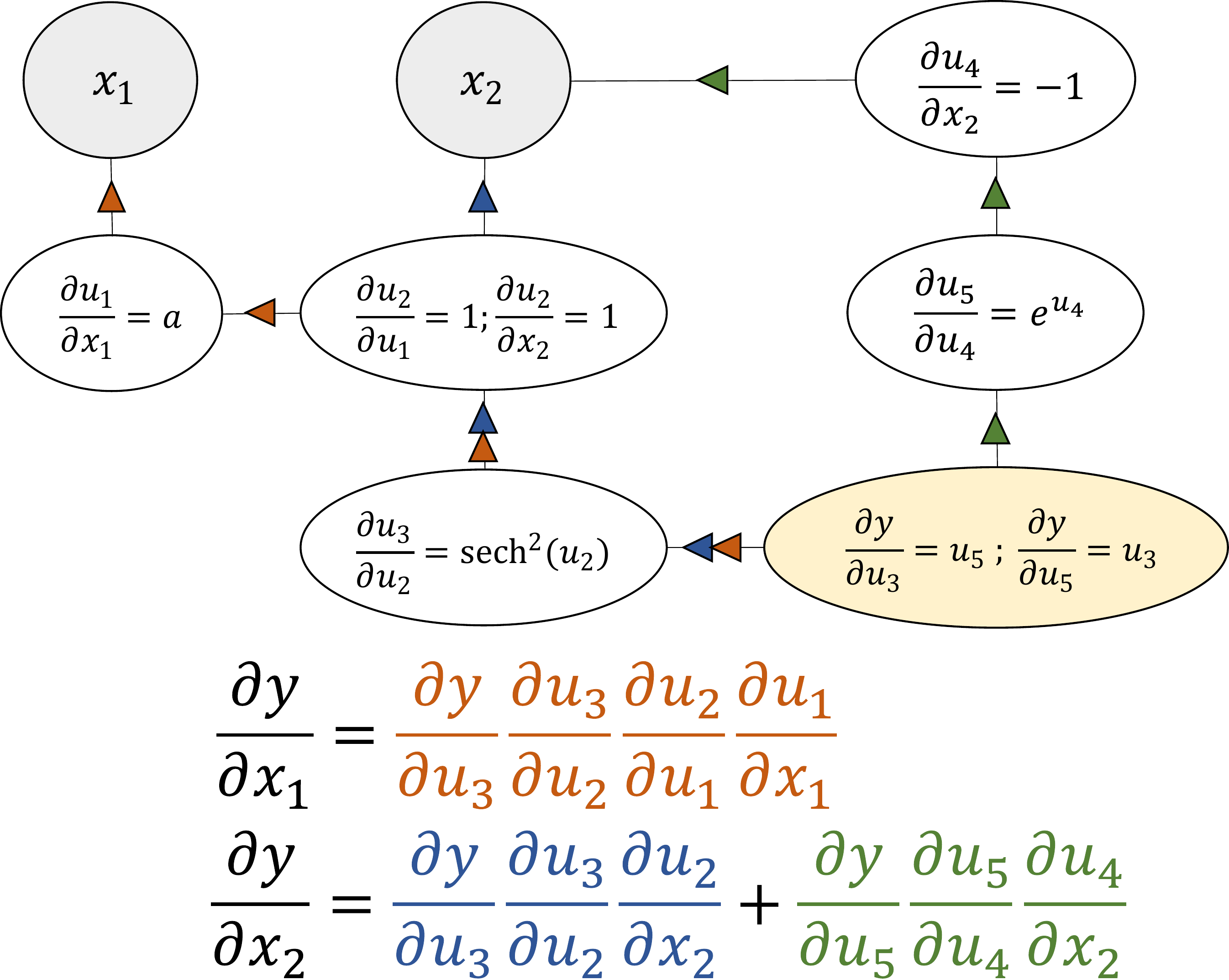}
\caption{Backwards propagation of gradients based on the chain rule through the graph constructed in Fig.~\ref{forward}.}
\label{backward}
\end{figure}

Backward mode AD, as implemented in PyTorch,\cite{pytorch} is used for this work. For illustration of the basic principle, consider evaluating the partial derivatives of $y(x_1, x_2) = \tanh{(ax_1 + x_2)} ~e^{-x_2}$. When computing $y(x_1, x_2)$, PyTorch constructs a graph tracking the elementary operations, as depicted in Fig.~\ref{forward}. Then, the derivatives $\partial y/\partial x_1$ and $\partial y/\partial x_2$ may be accumulated from elementary derivatives, in a manner consistent with the chain rule, during backwards traversal of the graph---see Fig.~\ref{backward}.

\subsubsection{\label{xitorch_exp} Differentiable Functionals}

AD may also be extended to complicated numerical algorithms. For example, PROFESS-AD makes frequent use of $\xi$-torch\cite{xitorch} (pronounced “sigh-torch”), a PyTorch-based library providing differentiable functionals, including differentiable initial value problem solvers and symmetric matrix eigensolvers. (Note that this use of "functional" is more generic than that used in the DFT context.) A crucial feature of $\xi$-torch is its use of analytical expressions for the backward pass, which is simpler than direct backpropagation through multi-step algorithms, many of which are iterative.\cite{dqc} The $\xi$-torch minimization functional is of particular relevance to our work. Here, "minimization functional" refers to an iterative, often gradient-based algorithm for solving problems of the form $\mathbf{x}^\text{min} = \operatorname*{argmin}_{\mathbf{x}\in\mathbb{R}^m} f_{\boldsymbol{\theta}}(\mathbf{x})$ by minimizing the function $f_{\boldsymbol{\theta}} : \mathbb{R}^m \to \mathbb{R}$, where $\boldsymbol{\theta}$ is a parameter vector. The ideal solution for $\mathbf{x}^\text{min}$ obeys the stationary condition
\begin{equation}
\label{xitorch_minimization_condition}
   \frac{\partial f_{\boldsymbol{\theta}}}{\partial x_i} \Bigg\vert_{\mathbf{x}^\text{min}} = 0 .
\end{equation} 
Suppose we have a known scalar function of the optimal argument vector, $g(\mathbf{x}^\text{min})$. It will depend on \(\theta\) implicitly through \(\mathbf{x}^\text{min}\) and might additionally have explicit \(\theta\) dependence. $\xi$-torch facilitates seamless computation of the derivative 
\begin{equation}
\label{minimization_derivative}
    \frac{\partial g(\mathbf{x}^\text{min})}{\partial \theta_k} = \frac{\partial g(\mathbf{x})}{\partial \theta_k}\Bigg\vert_{\mathbf{x}^\text{min}}  + \sum^m_{i=1} \frac{\partial g(\mathbf{x})}{\partial x_i}\Bigg\vert_{\mathbf{x}^\text{min}} \frac{\partial x_i^\text{min}}{\partial \theta_k},
\end{equation}
in a backward mode AD framework by determining the nontrivial $\partial \mathbf{x}^\text{min}/\partial \theta_k$ term analytically via
\begin{equation}
    \frac{\partial x^\text{min}_i}{\partial \theta_k} = - \sum_j \left[\left( \frac{\partial^2 f_{\boldsymbol{\theta}}}{\partial \mathbf{x}_1 \partial \mathbf{x}_2} \Bigg\vert_{\mathbf{x}^\text{min}} \right)^{-1}\right]_{ij} \frac{\partial^2 f_{\boldsymbol{\theta}}}{\partial \theta_k \partial x_j} \Bigg\vert_{\mathbf{x}^\text{min}},
\end{equation}
which can be derived by differentiating the minimization condition (Eq.~\ref{xitorch_minimization_condition}) with respect to $\theta_k$. Equivalently, the computation of $\partial \mathbf{x}^\text{min}/\partial \theta_k$ involves solving the matrix equation
\begin{equation}
    \sum_j \frac{\partial^2 f_{\boldsymbol{\theta}}}{\partial x_i \partial x_j} \Bigg\vert_{\mathbf{x}^\text{min}} \frac{\partial x^\text{min}_j}{\partial \theta_k} = - \frac{\partial^2 f_{\boldsymbol{\theta}}}{\partial \theta_k \partial x_i} \Bigg\vert_{\mathbf{x}^\text{min}}.
\end{equation}
As $\xi$-torch uses a differentiable linear equation solver for that purpose, whose backward pass is also differentiable, derivatives of arbitrarily high orders are accessible in principle. However, the accumulation of error and magnification of numerical noise during higher-order derivative computation complicates this task in practice.

\subsection{\label{density_optimization_scheme} Orbital-free Density Functional Theory}

The foundations of OFDFT are the Hohenberg-Kohn theorems,\cite{HK} which demonstrate that the ground state properties of a many-electron system may be found by minimizing an energy functional of the electron density $E[n]$, subject to the constraints that the density is nonnegative and integrates to the expected number of electrons, $N_e$. To enforce nonnegativity and normalization of the density, one may minimize with respect to an unconstrained variable $\chi(\mathbf{r})$, where
\begin{equation}
    n[\chi](\mathbf{r}) = \frac{N_e}{\tilde{N}} \chi^2(\mathbf{r}),
\end{equation}
with $\tilde{N} = \int \chi^2(\mathbf{r}')d^3\mathbf{r}'$. Effectively, this scheme minimizes $E[n[\chi]]$ with respect to $\chi(\mathbf{r})$, determining the ground state density, $n_\text{gs}$ from the optimized $\chi_\text{gs}$, which satisfies the minimization condition
\begin{equation}
\label{dEdchi_stationary}
    \frac{\delta E}{\delta \chi} \Bigg\vert_{\chi_\text{gs}} = 0.
\end{equation}
The relationship between $\delta E/\delta \chi$ and the more conventional density functional derivative $\delta E/\delta n$ is 
\begin{equation}
    \frac{\delta E}{\delta \chi(\mathbf{r})} 
    = 2 \chi(\mathbf{r}) \frac{N_e}{\tilde{N}} \left[ \frac{\delta E}{\delta n(\mathbf{r})} - \frac{1}{N_e} \int d^3\mathbf{r}' ~\frac{\delta E}{\delta n(\mathbf{r}')} n(\mathbf{r}') \right].
\end{equation}
This result is derived in the Supplementary Information. Among other benefits, recasting the constrained density optimization problem as a generic unconstrained minimization problem is useful for compatibility with $\xi$-torch's differentiable minimizer.\\

\begin{figure}[tb]
\centering
\includegraphics[width=\linewidth]{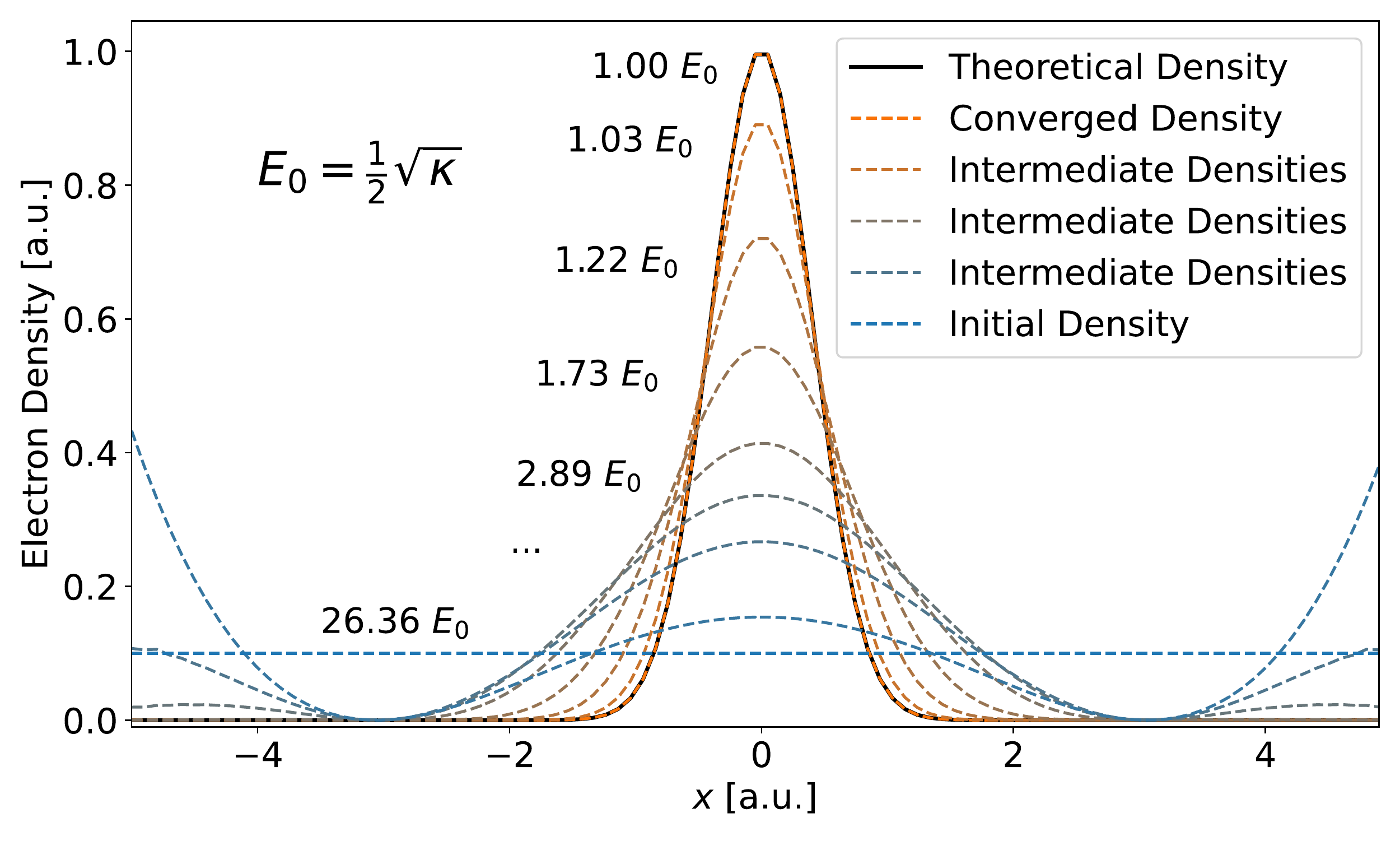}
\caption{Illustration of the OFDFT energy minimization procedure for a single electron in a one-dimensional harmonic oscillator potential with $\kappa=10$ (see Eq.~\ref{QHO_DFE}). The ground state energy of this system is $E_0  =(1/2)\sqrt{\kappa}$. The total energy corresponding to each density profile is written beside it as a factor of $E_0$, highlighting the steady decrease in energy as the system evolves to the expected ground state.}
\label{den_opt}
\end{figure}

To illustrate OFDFT energy minimizations, it is useful to consider a single, one-dimensional electron in a harmonic oscillator potential. An exact orbital-free description is available for this system because the Schr\"{o}dinger equation,
\begin{equation}
    \left(-\frac{1}{2} \frac{d^2}{dx^2} + \frac{1}{2} \kappa x^2 \right) \psi(x) = E\psi(x),
\end{equation}
may be converted into the density functional expression,
\begin{equation} \label{QHO_DFE}
    E[n] = \int_{-\infty}^\infty -\frac{1}{2} \sqrt{n(x)} \frac{d^2\sqrt{n(x)}}{dx^2} dx + \int_{-\infty}^\infty \frac{1}{2} \kappa x^2 n(x)  dx,
\end{equation}
using the relationship $\psi = \sqrt{n}$ between the single orbital $\psi$ and the electron density $n$. The variable $\kappa$ represents the curvature of the harmonic potential. Fig.~\ref{den_opt} depicts the evolution of the electron density and energy for this system over numerous gradient descent steps of the form $\chi_{i+1} = \chi_i - \gamma~\delta E/\delta \chi \vert_{\chi_i}$ from an initialized uniform density until convergence of the energy is achieved. ($\gamma$ controls the magnitude of the gradient descent steps).

\subsection{Automatic Differentiation and Orbital-free Density Functional Theory}

\subsubsection{Functional Derivatives}
Given a density functional $F[n]$, the functional derivative $\delta F/\delta n(\mathbf{r})$  is defined such that
\begin{equation}
    \int \frac{\delta F}{\delta n(\mathbf{r})} \phi(\mathbf{r}) d^3\mathbf{r} =  \frac{dF[n+\epsilon \phi]}{d \epsilon} \Bigg\vert_{\epsilon=0}.
\end{equation}

This quantity is crucial for OFDFT energy minimizations; conventionally, the functional derivative of each term contributing to the total energy functional must be derived and coded by hand. With AD, energy minimizations are achievable immediately---one need only implement the base functional, after which its (often complicated) functional derivative is determined automatically.\\

\subsubsection{\label{general_derivatives} Derivatives of the Ground State Energy}
With the unconstrained minimization from Section~\ref{density_optimization_scheme} in mind, consider an energy functional $E[\chi]$ that depends on an external parameter $\lambda$. Possible quantities represented by $\lambda$ include atom coordinates or lattice vector elements. This section investigates derivatives of the ground state energy, $E[\chi_\text{gs}]$, with respect to $\lambda$. The first such derivative is
\begin{equation}
\label{first-order}
    \frac{\partial E[\chi_\text{gs}]}{\partial \lambda} 
    = \frac{\partial E[\chi]}{\partial \lambda}\Bigg\vert_{\chi_\text{gs}} + \int d^3\mathbf{r}~ \underbrace{\frac{\delta E}{\delta \chi(\mathbf{r})}\Bigg\vert_{\chi_\text{gs}}}_{0} \frac{\partial \chi_\text{gs}(\mathbf{r})}{\partial \lambda}
\end{equation}
where the second term vanishes due to Eq.~\ref{dEdchi_stationary}. This result is a variant of the well-known Hellmann-Feynman theorem.\cite{hellmann, feynman} Its practical significance is that we may obtain first derivatives without knowledge of the challenging $\partial \chi_\text{gs}/\partial \lambda$ (or equivalently $\partial n_\text{gs}/\partial \lambda$) term. Accordingly, standard AD suffices. \\

For the second-order derivative, we differentiate the first derivative with respect to another parameter $\nu$ to obtain
\begin{widetext}
\begin{equation}
\begin{aligned}
\label{second-order}
    \frac{\partial^2 E[\chi_\text{gs}]}{\partial \lambda \partial \nu} 
    &= \frac{\partial^2 E[\chi]}{\partial \lambda \partial \nu}\Bigg\vert_{\chi_\text{gs}} + \int d^3\mathbf{r}~\left[\frac{\partial}{\partial \lambda} \left(\frac{\delta E}{\delta \chi(\mathbf{r})}\right) \right]\Bigg\vert_{\chi_\text{gs}} \frac{\partial \chi_\text{gs}(\mathbf{r})}{\partial \nu} \\
    &= \frac{\partial^2 E[\chi]}{\partial \lambda \partial \nu}\Bigg\vert_{\chi_\text{gs}} 
    + \int d^3\mathbf{r}\int d^3\mathbf{r}'~\left[\frac{\partial}{\partial \lambda} \left(\frac{\delta E}{\delta \chi(\mathbf{r})}\right) \right]\Bigg\vert_{\chi_\text{gs}} 
    \frac{\partial \chi_\text{gs}(\mathbf{r})}{\partial [\delta E/\delta \chi(\mathbf{r}')]\vert_{\chi_\text{gs}}} \left[\frac{\partial}{\partial \nu} \left(\frac{\delta E}{\delta \chi(\mathbf{r}')}\right) \right]\Bigg\vert_{\chi_\text{gs}}
\end{aligned} .
\end{equation}
\end{widetext}
Here, one can no longer avoid terms of the form $\partial \chi_\text{gs}/\partial \nu$. For this reason, such second derivatives are conventionally computed by finite differences (or, in some cases, with the aid of perturbation theory). One straightforward AD strategy (at least conceptually) would be direct backpropagation through the iterations, yielding $\chi_\text{gs}$. However, the analytical backward pass outlined in Section~\ref{xitorch_exp} offers a more direct route. To gain more insight into the underlying operations performed, one may differentiate Eq.~\ref{dEdchi_stationary} with respect to $\nu$ and rearrange to yield an expression for $\partial \chi_\text{gs}/\partial \nu$, which may be used to re-express Eq.~\ref{second-order} as

\begin{widetext}
\begin{equation}
\label{second-derivative_clean}
    \frac{\partial^2 E[\chi_\text{gs}]}{\partial \lambda \partial\nu} 
    = \frac{\partial^2 E[\chi]}{\partial \lambda \partial \nu}\Bigg\vert_{\chi_\text{gs}}  - \int d^3\mathbf{r}\int d^3\mathbf{r}'~\left[\frac{\partial}{\partial \lambda} \left(\frac{\delta E}{\delta \chi(\mathbf{r})} \right) \right]\Bigg\vert_{\chi_\text{gs}}
    \left( \frac{\delta^2 E}{\delta \chi(\mathbf{r}) \delta \chi(\mathbf{r}') }\Bigg\vert_{\chi_\text{gs}} \right)^{-1} \left[\frac{\partial}{\partial \nu} \left(\frac{\delta E}{\delta \chi(\mathbf{r}')}\right) \right]\Bigg\vert_{\chi_\text{gs}}.
\end{equation}
\end{widetext}

\begin{figure}[tb]
\centering
\includegraphics[width=\linewidth]{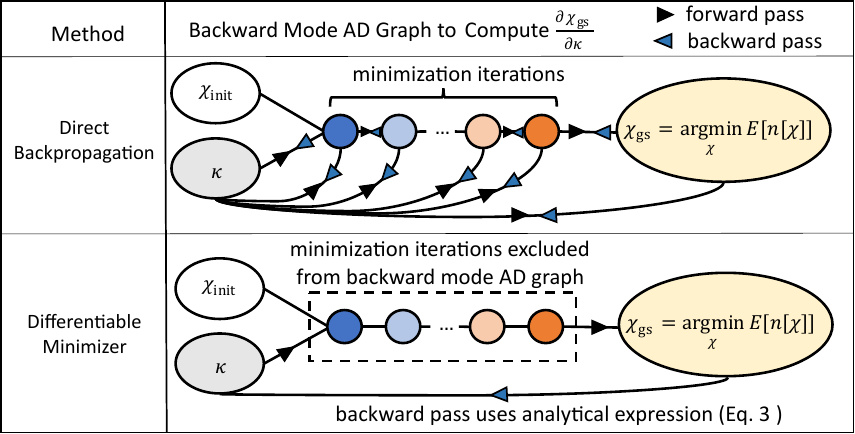}
\caption{Illustration of the differences between using direct backpropagation and $\xi$-torch's differentiable minimizer for computing the quantity $\delta \chi_\text{gs}/\delta \kappa$ for the single-electron one-dimensional quantum harmonic oscillator system.}
\label{xitorch_graph}
\end{figure}

To further illustrate these ideas, we return to the model system of a single-electron one-dimensional quantum harmonic oscillator. The parameter $\kappa$ determines the external potential and consequently affects the ground state energy and density. As special cases of Eqs.~\ref{first-order} and \ref{second-order}, the corresponding derivatives for this system are
\begin{equation}
    \frac{\partial E_\text{gs}}{\partial \kappa} = \int_{-\infty}^\infty dx \frac{1}{2} x^2 n[\chi_\text{gs}](x) 
\end{equation}
and 
\begin{equation}
    \frac{\partial^2 E_\text{gs}}{\partial \kappa^2} = \int_{-\infty}^\infty dx \frac{1}{2} x^2 \int_{-\infty}^\infty dx' \frac{\delta n(x)}{\delta \chi(x')}\bigg\vert_{\chi_\text{gs}} \frac{\partial \chi_\text{gs}(x')}{\partial \kappa}.
\end{equation}
As with the general case, the first $\kappa$ derivative of the ground state energy lacks any dependence on $\delta \chi_\text{gs}/\delta \kappa$ while the second $\kappa$ derivative does depend on this nontrivial term. As highlighted earlier, two distinct AD-based methods may be used for this term, direct backpropagation over the minimization iterations or the analytical approach outlined in Section~\ref{xitorch_exp}. The conceptual differences between the two methods are illustrated in Fig.~\ref{xitorch_graph}. Unlike direct backpropagation, the direct backward pass of $\xi$-torch's differentiable minimizer is independent of the minimization iterations and does not require caching of intermediate values from the minimization iterations, making it the more favourable option in most cases.

\section{PROFESS-AD: A Fully Differentiable OFDFT Code}
\label{PROFESS-AD_intro}

\subsection{Overview}

This section introduces PROFESS-AD, an auto-differentiable OFDFT code built using PyTorch. PROFESS-AD was designed as a prototyping tool to facilitate the development and testing of new methods, including new density functionals, with the expectation that features developed on this platform will eventually be implemented in mainstream performance-optimized codes, such as PROFESS.\cite{profess1,profess2,profess3} Several use cases for PROFESS-AD are listed next.
\begin{itemize}
    \item \textbf{Rapid prototyping}. A key indicator of the quality of a kinetic energy functional is its performance during the minimization procedure that yields the ground state energy and density. Conventional testing requires one to derive and implement the functional derivatives of a candidate functional. Furthermore, tests involving geometry optimizations require stresses to be derived and implemented, which becomes a genuine challenge when a functional is complicated. PROFESS-AD provides these derivatives immediately, accelerating the transition from concept to practical calculations.
    \item \textbf{Testing hand-coded derivatives}. At times, analytical derivatives are required, whether for implementation in mainstream codes or assessment of theoretical properties. PROFESS-AD's AD tools are useful for verifying the correctness of such derivative expressions and code.
    \item \textbf{Training highly-parameterized functionals}. A fully-differentiable OFDFT code facilitates the development of highly parameterized functionals by enabling parameter optimization via gradient-based methods. Only an energy functional expression is required for the fitting procedure. Moreover, for kinetic energy functionals, PROFESS-AD provides functions for computing conventional fitting targets such as the kinetic potential and the implied linear response function associated with a free electron gas. 
    \item \textbf{Higher-order derivatives}. PROFESS-AD streamlines the process of obtaining more complicated derivatives such as bulk moduli, elastic constants and force constants for more comprehensive benchmarking. This feature is especially useful given that external packages are often required for such calculations.
\end{itemize}

\subsection{Design Choices and Implementation Details}

\subsubsection{Programming Platform}
Use of PyTorch\cite{pytorch} is consistent with the features of Python that make it an attractive rapid-prototyping language, including easy-to-learn syntax, polymorphism, and object-oriented programming support.\cite{python_prototyping} (The OFDFT code DFTPy\cite{DFTPy} was written in pure Python for similar reasons.) An additional advantage is seamless GPU-compatibility, such that PROFESS-AD users can benefit from GPU-accelerated computation. \\

The choice of PyTorch is also consistent with the growing use of machine-learning techniques for density functional development, including machine-learned kinetic energy functionals for three-dimensional systems.\cite{yp, sergei2019, sergei2020, nakai2018, nakai2019, nakai2020, imoto, kumar} PROFESS-AD complements these efforts by simplifying incorporation of such machine-learning models in density functionals.

\subsubsection{Energy Functionals}

For a given configuration of atoms, the total energy functional is 
\begin{equation}
\label{total_energy}
\begin{aligned}
    E[n] 
    =  T_\text{S}&[n] + U_\text{Hartree}[n] + E_\text{XC}[n] \\
    &+ U_\text{ion-elec}[n] + U_\text{ion-ion}
\end{aligned}~~,
\end{equation}
where \(n\) is the electron density. $T_\text{S}$ is the noninteracting kinetic energy functional, $U_\text{ion-elec}$ is the ion-electron interaction energy, $U_\text{Hartree}$ is a classical mean-field electron-electron interaction energy, $E_\text{XC}$ is the exchange-correlation (XC) functional, and $U_\text{ion-ion}$ is the ion-ion electrostatic energy.\\

\textbf{Kinetic Energy} \\

Noninteracting kinetic energy functionals can be broadly categorized as semilocal functionals or nonlocal functionals. \\

Semilocal functionals involve a single integral of an energy density, which at a given point may depend on the electron density and related quantities like the gradient or Laplacian of the density at that point. Two fundamental semilocal functionals yield the Weizs\"{a}cker\cite{vW} and Thomas-Fermi\cite{Thomas,Fermi} kinetic energies, with the former exact for single-orbital systems and the latter exact for the homogeneous electron gas. Additional semilocal functionals available at present include the vWGTF,\cite{vWGTF} Luo-Karasiev-Trickey,\cite{LKT} and Pauli-Gaussian\cite{PG1,PG2} functionals.\\

Nonlocal functionals have more complicated functional forms, generally involving at least two integrals. One well-known class of functionals suitable for nearly-free-electron metals was pioneered by Wang and Teter, which includes the Wang-Teter,\cite{wt} Perrot,\cite{perrot} Smargiassi-Madden,\cite{sm} and density-independent Wang-Govind-Carter\cite{wgc98,wgc99} functionals. A generalization of the standard Wang-Teter functional incorporates the quadratic response of the free-electron gas in addition to the linear response;\cite{wt,fm,response1,response2} PROFESS-AD implements the Foley-Madden\cite{fm} version of this latter functional. The Wang-Teter class has also been adapted for random structure searching by introduction of  a Pauli-stabilization function ensuring the positivity of the Pauli kinetic energy.\cite{ofdft_rss} More significant departures include the density-dependent Wang-Govind-Carter\cite{wgc99,wgc99_analytical} functional via the introduction of a density-dependent kernel. To extend the application of OFDFT to gapped materials, the Huang-Carter\cite{hc} functional was proposed, with its numerical stability issues addressed by the revised Huang-Carter\cite{revHC} functional. For these, PROFESS-AD employs an interpolation scheme similar to that used in Ref.~\onlinecite{hc} to make the kernel convolution scale as $O(N \log N)$. An alternative for the description of gapped materials is the KGAP functional,\cite{kgap} which uses a similar Wang-Teter style decomposition of energy terms but imposes the jellium-with-gap linear response\cite{jgap1,jgap2} instead of the Lindhard response. Finally, PROFESS-AD offers a few functionals with kernels derived from functional integration, including  the Mi-Genova-Pavanello\cite{mgp} and Xu-Wang-Ma\cite{xwm} functionals.\\

\textbf{Electrostatic Energies}\\

For the ion-electron term, users can choose between a direct quadratic-scaling implementation of the structure factor or a $O(N\log N)$ scaling approximation based on the particle-mesh Ewald (PME) scheme.\cite{pme1,pme2,pme3} The Hartree energy is computed via FFTs. Often, the ion-ion interaction term $U_\text{ion-ion}$ is computed by Ewald summation;\cite{ewald1,ewald2,ewald3} however PROFESS-AD utilizes the real-space method of Pickard\cite{ion-ion} for its simplicity.\\

\textbf{Exchange-Correlation Energy}\\

At present, PROFESS-AD offers local density approximation (LDA) XC functionals based on both the Perdew-Zunger\cite{PZ} and Perdew-Wang\cite{PW92} parameterizations of the Ceperley-Alder quantum Monte Carlo data for the homogeneous electron gas.\cite{CA} Additionally, the nonempirical Chachiyo LDA correlation, derived from second-order M{{\o}}ller-Plesset perturbation theory, is implemented.\cite{chachiyo} At the generalized gradient approximation (GGA) level, the Perdew-Burke-Ernzerhof (PBE) functional is available.\cite{PBE}\\

\subsubsection{OFDFT Scheme}

The energy minimization procedure presented in Section~\ref{density_optimization_scheme} is compatible with any gradient-based minimization algorithm. PROFESS-AD offers the choice between a modified (PyTorch-based) limited-memory Broyden–Fletcher–Goldfarb–Shanno (LBFGS) optimizer\cite{lbfgs_torch} and a two-point gradient descent algorithm.\cite{2pgd}

\section{\label{PROFESS-AD_applications} Features and Applications}

This section presents example applications. All OFDFT calculations were performed with PROFESS-AD using a plane-wave cutoff of $2000$ \si{eV} for the electron density and bulk-derived local pseudopotentials (LPPs).\cite{BLPS} For comparison, KSDFT calculations were performed with CASTEP,\cite{castep} using a plane-wave cutoff of $1000$ \si{eV} for the wavefunctions and Brillouin zone sampling with Monkhorst-Pack\cite{mp_grid} grids having maximum distance between k-points of $0.015 \times 2\pi$ \si{\angstrom^{-1}}. KSDFT calculations were performed with the same LPPs
used for the OFDFT calculations, as well as the C19 set of ultrasoft nonlocal pseudopotentials (NLPPs). The KSDFT-LPP results allow for assessment of the accuracy of the approximate kinetic energy functionals used in OFDFT calculations, while the KSDFT-NLPP calculations provide additional information about the suitability of the LPPs. In all calculations, the PBE\cite{PBE} XC functional was used.\\

Our test systems comprise the face-centered cubic (fcc), hexagonal close packed (hcp), body-centered cubic (bcc), simple cubic (sc) and diamond cubic (dc) structures of the elemental metals Li, Mg, and Al, as well as the wurtzite GaAs system.

\subsection{Functional Derivatives for Energy Minimization}

The functional derivatives required for energy minimization are determined effortlessly with AD, accelerating the functional development process---one can proceed directly from a new functional form to benchmarking.\\

To illustrate this feature, we performed standard equation-of-state tests with several kinetic energy functionals, including a published functional (the Yuk1 functional\cite{yukawa1}) that has not yet been assessed in literature for its performance in energy minimizations (to the knowledge of the authors). The Yukawa kinetic energy functionals\cite{yukawa1,yukawa2} involve a nonlocal ingredient based on the Yukawa potential, 
\begin{equation}
    y_{\alpha \beta} (\mathbf{r}) 
    = \frac{3\pi\alpha^2}{4 k_F(\mathbf{r}) n^{\beta-1}(\mathbf{r})} \int d^3\mathbf{r}' \frac{n^\beta(\mathbf{r}') e^{-\alpha k_F(\mathbf{r}) |\mathbf{r}-\mathbf{r}'|}}{|\mathbf{r}-\mathbf{r}'|},
\end{equation}
where $k_F(\mathbf{r}) = [3\pi^2 n(\mathbf{r})]^{1/3}$ is the spatially varying Fermi wavevector. These functionals were developed to respect exact constraints such as the Lindhard constraint\cite{lindhard_exp} and Pauli positivity\cite{levy_ouyang} as best they can. However, while they were tested on converged Kohn-Sham densities for spherical systems, their performance in energy minimzations and for periodic systems is mostly unknown, having been mentioned as avenues for future work.\cite{yukawa1} \\

Our tests consist of equilibrium equation-of-state (EOS) fits for the zero-pressure properties of various simple structures (fcc, hcp, bcc, sc, dc) of Li, Mg, and Al. We employed the Yuk1 functional,\cite{yukawa1} as well as the PGSLr\cite{PG2} and Wang-Teter (WT)\cite{wt} functionals for comparison. Energy-volume curves were constructed with 11 evenly spaced points within $\pm 5\%$ of its equilibrium volume, and fit with the Birch-Murnaghan EOS.\cite{bm_eos} To compute the Yukawa descriptor, $y_{\alpha\beta}$ efficiently, we used a similar interpolation scheme as that used for the Huang-Carter functionals. Table~\ref{yuk_benchmark} presents the results of these calculations, with reference KSDFT-LPP calculations from Ref.~\onlinecite{ofdft_rss} included for comparison.\\

It is clear from the tables that the WT functional performs best overall in the description of these simple metals, likely because it obeys the Lindhard constraint exactly while the Yuk1 and PGSLr functionals only obey it approximately. As the Lindhard constraint is based on perturbations on the free electron gas, it is particularly relevant to nearly free electron like metallic systems. Of course, there are other considerations for comparing these functionals; nevertheless these calculations highlight the utility of PROFESS-AD for rapid prototyping. Further work could consider subsequent generations of the Yukawa functionals, such as Yuk3.\cite{yukawa1,yukawa2}

\begin{table}[htb]
\centering
\caption{Comparison of the Yuk1, PGSLr and WT functionals in terms of their description of the simple structures of elemental Li, Mg, and Al. $V_0$ is the equilibrium volume (\si{\angstrom^3} per atom), $\Delta E_0$ is the relative equilibrium energy (\si{eV} per atom) and $K_0$ is the equilibrium bulk modulus (\si{GPa}). The OFDFT results numerically closest to the KSDFT results are \textbf{bolded}.}
\begin{tabular}{lcccccc}
\hline \hline
\multicolumn{1}{c}{Li}           & \multicolumn{1}{l}{} & fcc           & hcp           & bcc           & sc            & dc            \\ \hline
\multicolumn{1}{c}{$V_0$}        
                                 & KS-LPP               & 20.2          & 20.2          & 20.2          & 21.9          & 30.5          \\
                                 & OF-Yuk1              & 20.9          & 20.9          & 20.9          & 22.5          & 27.5          \\
                                 & OF-PGSLr             & 19.9          & 19.9          & 19.8          & 22.1          & 28.9          \\
                                 & OF-WT                & \textbf{20.2} & \textbf{20.3} & \textbf{20.2} & \textbf{21.9} & \textbf{30.3} \\ \hline
\multicolumn{1}{c}{$\Delta E_0$} 
                                 & KS-LPP               & 0             & 0             & 1             & 152           & 428           \\
                                 & OF-Yuk1              & 0             & -1            & -1            & 139           & 450           \\
                                 & OF-PGSLr             & 0             & \textbf{0}    & 4             & \textbf{151}  & 423           \\
                                 & OF-WT                & 0             & \textbf{0}    & \textbf{1}    & \textbf{151}  & \textbf{429}  \\ \hline
\multicolumn{1}{c}{$K_0$}       
                                 & KS-LPP               & 16            & 16            & 16            & 12            & 6             \\
                                 & OF-Yuk1              & 15            & 15            & 15            & 13            & 9             \\
                                 & OF-PGSLr             & 17            & 17            & 16            & \textbf{12}   & 8             \\
                                 & OF-WT                & \textbf{16}   & \textbf{16}   & \textbf{16}   & \textbf{12}   & \textbf{6}    \\ \hline \hline

\multicolumn{1}{c}{Mg}           & \multicolumn{1}{l}{} & hcp           & fcc           & bcc           & sc            & dc            \\ \hline
\multicolumn{1}{c}{$V_0$}       
                                 & KS-LPP               & 22.9          & 23.1          & 22.8          & 27.1          & 39.9          \\
                                 & OF-Yuk1              & 24.6          & 24.6          & 24.5          & \textbf{27.0} & 35.4          \\
                                 & OF-PGSLr             & 24.3          & 24.2          & 24.3          & \textbf{27.2} & 35.5          \\
                                 & OF-WT                & \textbf{23.1} & \textbf{23.2} & \textbf{23.0} & \textbf{27.2} & \textbf{39.5} \\ \hline
\multicolumn{1}{c}{$\Delta E_0$} 
                                 & KS-LPP               & 0             & 14            & 29            & 410           & 854           \\
                                 & OF-Yuk1              & 0             & 0             & 8             & \textbf{396}  & 1143          \\
                                 & OF-PGSLr             & 0             & 0             & 11            & 292           & 831           \\
                                 & OF-WT                & 0             & \textbf{10}   & \textbf{26}   & 391           & \textbf{839}  \\ \hline
\multicolumn{1}{c}{$K_0$}       
                                 & KS-LPP               & 39            & 38            & 38            & 24            & 10            \\
                                 & OF-Yuk1              & \textbf{37}   & \textbf{37}   & \textbf{37}   & 29            & 17            \\
                                 & OF-PGSLr             & 33            & 35            & 32            & 25            & 17            \\
                                 & OF-WT                & \textbf{37}   & \textbf{37}   & \textbf{37}   & \textbf{24}   & \textbf{11}   \\ 
\hline \hline
\multicolumn{1}{c}{Al}           & \multicolumn{1}{l}{} & fcc           & hcp           & bcc           & sc            & dc            \\ \hline
\multicolumn{1}{c}{$V_0$}       
                                 & KS-LPP               & 16.6          & 16.7          & 17            & 19.9          & 27.3          \\
                                 & OF-Yuk1              & 17.3          & 17.3          & 17.3          & 19.5          & \textbf{26.3} \\
                                 & OF-PGSLr             & 18.1          & 18.1          & 18.2          & 20.4          & 26.2          \\
                                 & OF-WT                & \textbf{16.8} & \textbf{16.9} & \textbf{17.2} & \textbf{19.9} & 28.8          \\ \hline
\multicolumn{1}{c}{$\Delta E_0$} 
                                 & KS-LPP               & 0             & 25            & 80            & 335           & 723           \\
                                 & OF-Yuk1              & 0             & 1             & 27            & 670           & 1835          \\
                                 & OF-PGSLr             & 0             & 1             & 24            & 362           & 1118          \\
                                 & OF-WT                & 0             & \textbf{18}   & \textbf{73}   & \textbf{312}  & \textbf{791}  \\ \hline
\multicolumn{1}{c}{$K_0$}        
                                 & KS-LPP               & 77            & 75            & 71            & 58            & 39            \\
                                 & OF-Yuk1              & 97            & 97            & 96            & 77            & 45            \\
                                 & OF-PGSLr             & 70            & 69            & 69            & 61            & \textbf{44}   \\
                                 & OF-WT                & \textbf{79}   & \textbf{77}   & \textbf{72}   & \textbf{58}   & 24   \\  \hline \hline
                                 
\end{tabular}
\label{yuk_benchmark}
\end{table}

\subsection{First Derivatives of the Ground State Energy}

PROFESS-AD uses AD to compute properties related to first-order derivatives that would conventionally be implemented manually. Such quantities include the pressure, $P = - \partial E_\text{gs}/\partial \Omega$ for cell colume $\Omega$, the stress, $\sigma_{ij} = (1/\Omega) \partial E_\text{gs}/\partial \epsilon_{ij}\vert_{\epsilon_{ij} = 0}$ for strain $\epsilon_{ij}$, and the force, $\mathbf{F}_{\kappa} = - \nabla_{\mathbf{R}_\kappa} E_\text{gs}$, where $\mathbf{R}_\kappa$ is the ionic coordinate of the ion indexed by $\kappa$. These quantities are relevant for geometry optimizations, which minimize the ground state energy with respect to the lattice vectors and ionic coordinates. Additionally, PROFESS-AD provides direct access to the the derivatives $\partial E_\text{gs}/\partial h_{ij}$, where $h_{ij}$ are elements of the matrix whose columns are the lattice vectors.\\

Furthermore, PROFESS-AD facilitates constrained geometry optimizations, where the lattice vectors and ion coordinates depend on a separate set of parameters. This feature is demonstrated by structure optimization of hcp Mg and wurtzite GaAs.

\begin{figure}[htb]
\centering
\includegraphics[width=\linewidth]{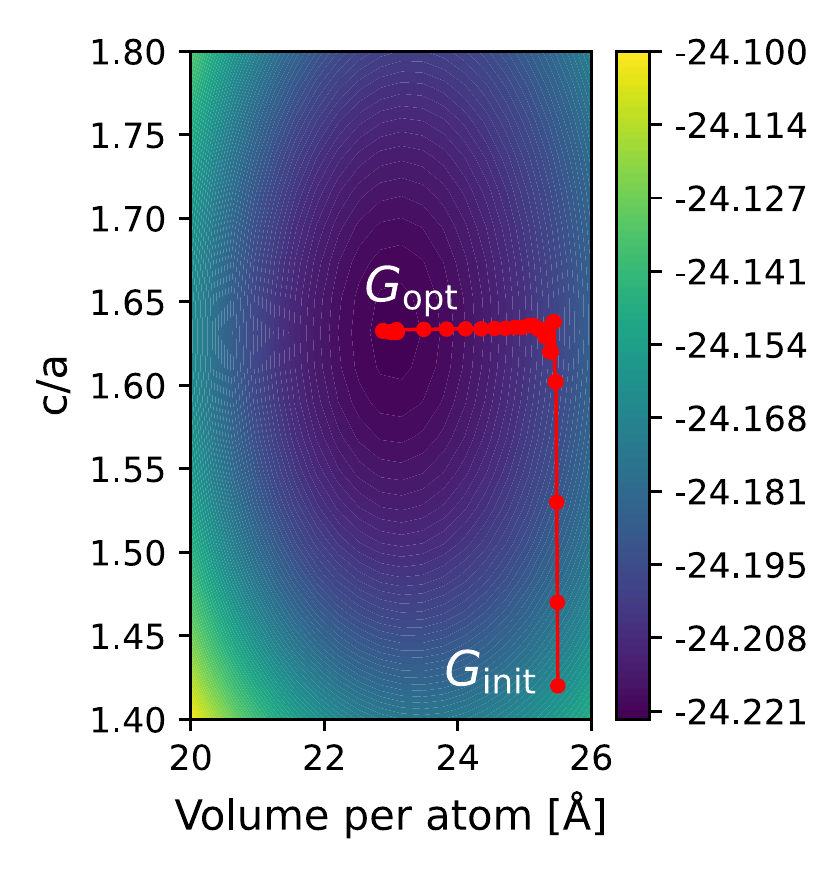}
\caption{Illustration of the parameterized geometry optimization of hcp-Mg. The colour bar on the right represents the total energy per atom in units of \si{eV}.}
\label{geom_opt}
\end{figure}

For hcp Mg, the ground state energy is minimized with respect to the $c/a$ ratio and cell volume, with the corresponding derivatives obtained directly with AD. The RPROP algorithm\cite{rprop} and the convergence condition that the maximum force and stress component must be below $10^{-4}$ \si{eV/\angstrom} and $10^{-4}$ \si{eV/\angstrom^3} respectively were used for this example. Fig.~\ref{geom_opt} depicts the optimization path from the initial geometry $G_\text{init}$ of volume = 25.50 \si{\angstrom} per atom and $c/a$ = 1.42 to the optimized geometry $G_\text{opt}$ of volume = 23.04  \si{\angstrom} per atom and $c/a$ = 1.63.

The second example involves wurtzite GaAs, where the energy is minimized with respect to its three degrees of freedom, the volume, the $c/a$ ratio, and an additional fractional coordinate parameter, $u$. The Huang-Carter functional\cite{hc} (with the parameters $\lambda=0.01177$ and $\beta=0.7143$) and revised Huang-Carter functional\cite{revHC} (with the parameters $a = 0.45$, $b = 0.10$ and $\beta = 2/3$) were used for this calculation. Forces and stresses were converged to be below $5 \times 10^{-2}$ \si{eV/\angstrom} and $6 \times 10^{-4}$ \si{eV/\angstrom^3} respectively for both OFDFT and KSDFT. The resulting structural parameters from OFDFT and KSDFT calculations, as well as experiment, are presented in Table~\ref{GaAs_table}. While reasonable agreement is observed, the errors between the OFDFT, KSDFT-LPP, and KSDFT-NLPP equilibrium volumes highlight the limitations of the kinetic energy functionals and local pseudopotentials for this system.

\begin{table}[htb]
\centering
\caption{Structure of wurtzite GaAs as predicted by OFDFT with the Huang-Carter (HC) and revised Huang-Carter (revHC) functionals, as well as KSDFT with local and nonlocal pseudopotentials (KSDFT-LPP and KSDFT-NLPP, respectively), and as observed in experiment. The relevant structural parameters are $V_\text{at}$ (the volume in units of \si{\angstrom^3} per atom), the $c/a$ ratio, and the fractional coordinate $u$.}
\begin{tabular}{ccccccc}
\hline \hline
              & $V_\text{at}$ & $c/a$ & $u$ \\ \hline
OFDFT-HC      & 22.602 & 1.657 & 0.3723     \\
OFDFT-revHC   & 22.658 & 1.651 & 0.3731     \\ 
KSDFT-LPP     & 23.109 & 1.650 & 0.3740      \\
KSDFT-NLPP    & 23.691 & 1.649 & 0.3740  \\ \hline
Experiment (Ref.~\onlinecite{wurtzite_GaAs})  & 22.615 & 1.646 & 0.3731    \\
\hline \hline
\end{tabular}
\label{GaAs_table}
\end{table}

\subsection{Second Derivatives of the Ground State Energy}

\subsubsection{Phonon Calculations with Ionic Coordinate Derivatives}
Within the harmonic approximation, the calculation of phonon properties depends on the second order force constants
\begin{equation}
\label{force_constants}
    \Phi_{ij}(\ell \kappa, \ell' \kappa') = \frac{\partial^2 E_\text{gs}}{\partial  u^{}_i(\ell \kappa) \partial u^{}_j(\ell'\kappa')},
\end{equation}
where $u_i(\ell \kappa)$ is an atomic displacement, $i, j \in \{x,y,z\}$ represent the Cartesian directions, $\ell, \ell'$ are unit cell labels, and $\kappa, \kappa'$ index the atoms in each unit cell. The force constants are used to construct the dynamical matrix, which is diagonalized to yield a phonon band structure that can be post-processed to estimate thermal properties.\\

Conventionally, such force constants are computed via finite difference approximations, where forces are obtained from separate calculations for a set of perturbations of the ionic positions. PROFESS-AD provides an exact method for determining force constants, whereby one need not worry about the magnitude of the finite difference step. \\

\begin{figure}[tb]
\centering
%\includesvg[scale=1.05]{bcc-li_phonon-dispersion}
\includegraphics[width=\linewidth]{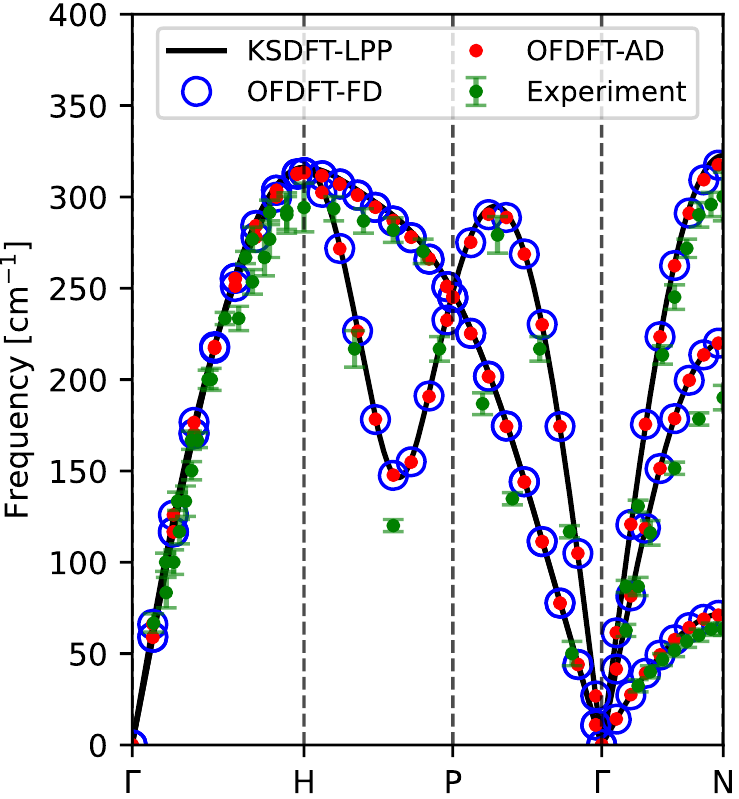}
\caption{Zero pressure phonon band structure for bcc-Li computed using Kohn-Sham (KS) and orbital-free (OF) DFT. The force constants from the OFDFT calculations were obtained using AD and finite differences (FD). Experimental data from Ref.~\onlinecite{phonon_expt} is presented for reference.}
\label{bcc-li_phonon-dispersion}
\end{figure}

As an example, the phonon band structure for bcc Li was generated using force constants computed with both finite differences and AD with the Wang-Teter kinetic functional.\cite{wt} These calculations were facilitated by Phonopy,\cite{phonopy} which generated the supercells required for both methods, and the supercells with displacements necessary for the finite difference calculations. For comparison, similar phonon calculations were performed with KSDFT using the nondiagonal supercell method.\cite{nondiagonal_supercells} In all calculations, the phonon band structures were converged to within 5 \si{cm^{-1}} when a $5 \times 5 \times 5$ supercell was used. The phonon band structures obtained from these calculations are presented in Fig.~\ref{bcc-li_phonon-dispersion}.\\

The agreement between the OFDFT band structures obtained from both finite differences and AD confirms that the two methods  yield the same force constants. Furthermore, the agreement between the KSDFT-LPP and OFDFT band structures indicates that the Wang-Teter functional can describe bcc-Li quite well, while deviation from the experimental data likely conveys limitations in the local pseudopotentials or the XC functional used.

\subsubsection{Elastic Properties with Lattice Vector Derivatives}

The bulk modulus, $K$, requires the second volume derivative of the ground state energy,
\begin{equation}
    K = \Omega \frac{\partial^2 E_\text{gs}}{\partial \Omega^2}.
\end{equation}
Furthermore, the second-order elastic constants, specifically the Birch coefficients,\cite{bm_eos,wallace} are defined as derivatives of the Cauchy stress $\sigma_{ij}$ with respect to the infinitesimal Cauchy strains $\epsilon_{k\ell}$,
\begin{equation}
\label{elastic_constants}
    C_{ijk\ell} = \frac{\partial \sigma_{ij}}{\partial \epsilon_{k\ell}}\Bigg\vert_{\epsilon_{ij} = 0}  = \frac{1}{2} \sum_m \left( \frac{\partial \sigma_{ij}}{\partial h_{km}} h_{\ell m} + \frac{\partial \sigma_{ij}}{\partial h_{\ell m}} h_{k m} \right),
\end{equation}
where the Cauchy stress tensor is
\begin{equation}
\label{stress}
    \sigma_{ij} = \frac{1}{\Omega} \frac{\partial E_\text{gs}}{\partial \epsilon_{ij}}\Bigg\vert_{\epsilon_{ij} = 0} = \frac{1}{2\Omega} \sum_k \left(\frac{\partial E_\text{gs}}{\partial h_{ik}} h_{jk} + \frac{\partial E_\text{gs}}{\partial h_{jk}} h_{ik} \right) .
\end{equation}
This definition of $C_{ijk\ell}$ only displays $i\leftrightarrow j$ and $k \leftrightarrow \ell$ symmetry but does not possess complete Voigt symmetry since $C_{ijk\ell} \neq C_{k\ell ij}$ in general. Under conditions of an arbitrary initial isotropic pressure however, the $C_{ijk\ell}$ coefficients do exhibit complete Voigt symmetry, and are the coefficients measured in wave propagation experiments for materials under isotropic pressure.\cite{wallace} The conversion of the stress and elastic constants from strain derivatives to lattice vector derivatives is presented in the Supplementary Information. \\

PROFESS-AD provides methods for computing the bulk modulus and elastic constants based on these definitions. The procedure is as simple as calling \texttt{system.elastic\_constants()} as AD provides these quantities directly, in contrast to more complicated multi-step procedures performed ordinarily.\cite{elastic_package_1, elastic_package_2} PROFESS-AD also provides tools for post-processing the elastic constants into other elastic quantities including the Reuss and Voigt bulk modulus and shear modulus, as well as Young's modulus and Poisson's ratio.\\

As an illustration, we investigate the elastic constants of fcc Al using the Foley-Madden (FM)\cite{fm} kinetic energy functional, whose complicated and lengthy functional form has historically rendered its stress contribution intractable. We compare the elastic constants obtained via AD to the more conventional stress-strain method, where the elastic constants are obtained by fitting the stress-strain data to the generalized Hooke's Law, $\sigma_{ij} = C_{ijk\ell} \epsilon_{k\ell}$. Specifically, nine stress-strain data points were evaluated (with the central data point corresponding to the unstrained structure), then fit with a quadratic polynomial from which the linear coefficient was extracted as the elastic constant. For comparison, KSDFT calculations were also performed with the stress-strain method.\\

\begin{figure}[tb]
\centering
%\includesvg[scale=1.05]{fcc-al_elastic-constants}
\includegraphics[width=\linewidth]{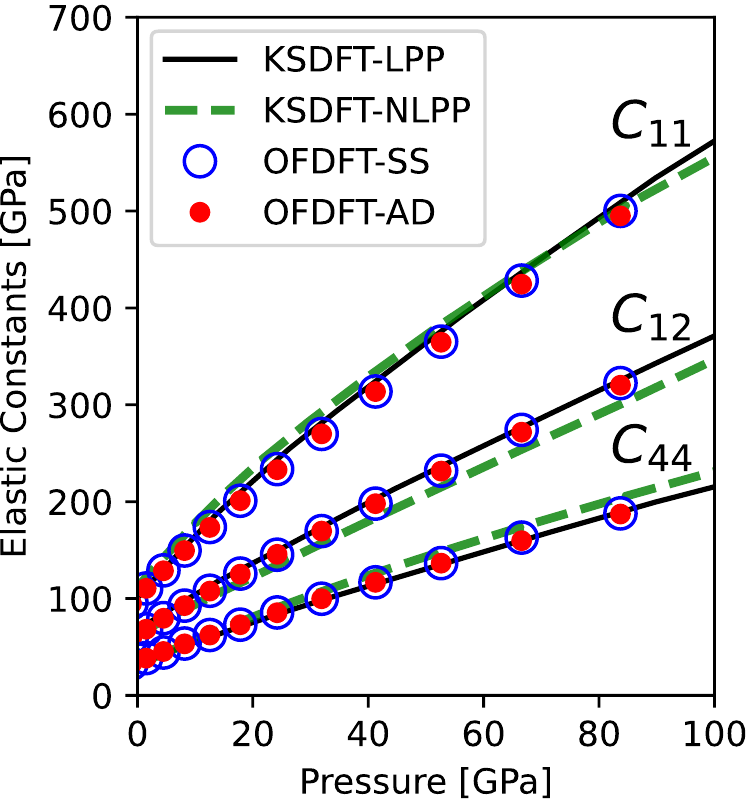}
\caption{High pressure elastic constants of fcc-Al at 0 \si{\kelvin} computed via orbital-free (OF) and Kohn-Sham (KS) DFT. For the OFDFT calculations, the elastic constants obtained via the stress-strain method (SS) and AD are compared.}
\label{fcc-al_elastic-constants}
\end{figure}

The zero pressure elastic property predictions are presented in Table~\ref{fcc-al_elastic-constants_0p}, along with experimental data, while Fig.~\ref{fcc-al_elastic-constants} presents OFDFT and KSDFT data at pressures up to 100 \si{GPa}. These results verify that the OFDFT elastic constants obtained with AD and the stress-strain method are practically identical, validating the much simpler AD method. We also note that the OFDFT predictions for fcc Al's elastic constants agree well with the KSDFT predictions.

\begin{table}[tb]
\centering
\caption{Zero-pressure elastic constants $C_{ij}$, bulk moduli $K$, shear moduli $G$, and Young's moduli $E$ of fcc-Al from theoretical calculations and experimental measurements. All elastic constants and moduli are presented in units of \si{GPa}.}
\begin{tabular}{ccccccc}
\hline \hline
              & $C_{11}$ & $C_{12}$ & $C_{44}$ & $K$ & $G$ & $E$ \\ \hline
OFDFT-AD      & 101 & 62 & 35 & 75 & 28 & 74      \\
OFDFT-SS      & 101 & 63 & 35 & 76 & 28 & 74      \\
KSDFT-LPP-SS     & 100 & 67 & 32 & 78 & 25 & 67      \\
KSDFT-NLPP-SS    & 111 & 63 & 33 & 79 & 29 & 78      \\ \hline
Experiment (Ref.~\onlinecite{fcc_al_expt1})  & 123 & 71 & 31 & 88 & 29  & 78      \\
Experiment (Ref.~\onlinecite{fcc_al_expt2})  & 114 & 62 & 32 & 79 & 29  & 79      \\
\hline \hline
\end{tabular}
\label{fcc-al_elastic-constants_0p}
\end{table}

\subsection{Tools for Functional Development}

PROFESS-AD provides several AD-based tools for computing additional derivatives of energy functionals. One example is a function for determining the quantity
\begin{equation} \label{lindhard}
    G^{-1}(\eta) = \frac{\pi^2}{k_F} \left( \hat{\mathcal{F}} \left\{ \frac{\delta^2 T_\text{S}}{\delta n(\mathbf{r}) \delta n(\mathbf{r}')} \right\} \Bigg\vert_{n_0}  \right)^{-1} ,
\end{equation}
where $n_0$ is a uniform density, $k_F = (3 \pi^2 n_0)^{1/3}$ is the Fermi wavevector, and $\eta = |\mathbf{q}|/2k_F$ is a dimensionless wavevector. The relevance of this quantity derives from a known constraint on the noninteracting kinetic energy functional, namely that $G^{-1}(\eta)$ should equal the Lindhard linear response function for a free electron gas,\cite{lindhard} given by
\begin{equation}
    G_\text{Lind}^{-1}(\eta) = \frac{1}{2} + \frac{1-\eta^2}{4\eta} \ln{\left|\frac{1+\eta}{1-\eta}\right|}.
\end{equation}

These AD-based tools are useful for verifying analytically derived derivatives and/or for debugging hand-coded derivatives. They also simplify development of parameterized functionals via gradient-based optimization, as their outputs are themselves auto-differentiable, allowing them to become training targets. For example, the kinetic potential is an increasingly popular loss function ingredient target for machine-learned functional developers.\cite{meyer, imoto} An example based on a mini-problem faced by Imoto et al.\cite{imoto} in the development of their machine-learned semilocal kinetic energy functional is presented in the Supplementary Information, which uses $G^{-1}(\eta)$ as the training target. Furthermore, PROFESS-AD trivially allows for less-traditional fitting targets such as the kinetic stress to be included in the loss function, and the availability of nontrivial ground state density derivatives $\partial n_\text{gs}/\partial \lambda$, enabled by $\xi$-torch, opens the door to even more sophisticated training procedures which could include fitting for the ground state densities and other ground state properties after a density optimization.

\section{Discussion}

The advantages AD provides do not come completely for free; the extra machinery does incur some overhead. However, in simple tests on fcc Al supercells with up to 1000 atoms, we compared the time required for AD-based functional differentiation against that required for manually coded derivatives, observing an additional cost never greater than twenty percent (and typically less).\\

Future work could improve the performance of PROFESS-AD, making it competitive even for large-scale performance-intensive applications. One straightforward avenue would be just-in-time compilation, but there are algorithmic aspects that could be optimized in tandem. For example, the present implementation of $\xi$-torch's differentiable minimizer does not utilize the knowledge that certain terms will vanish, such as in Eq.~\ref{first-order}.\\

\section{Conclusion}

This work introduced AD to the field of OFDFT in the form of the PROFESS-AD code. As illustrated by numerous examples, PROFESS-AD is suitable for practical calculations. In addition, it is especially promising as a prototyping tool for prompt and convenient testing. Additionally, we expect the ability to differentiate with respect to arbitrary external parameters will expand the reach of OFDFT into new application areas. Work of this kind is ongoing.

\section{Supplementary Material}

See Supplementary Material for the expanded derivations and examples mentioned above.

\begin{acknowledgments}
The authors thank Muhammad Kasim, Sam Vinko and Michael Herbst for helpful discussions.

W.C.W. was supported by the Schmidt Science Fellows in partnership with the Rhodes Trust. W.C.W. and C.J.P. acknowledge support from the EPSRC (Grant EP/V062654/1) and C.J.P. further acknowledges EPSRC support for the UKCP consortium (Grant EP/P022596/1).

This work was performed using resources provided by the Cambridge Service for Data Driven Discovery (CSD3) operated by the University of Cambridge Research Computing Service (www.csd3.cam.ac.uk), provided by Dell EMC and Intel using Tier-2 funding from the EPSRC (capital grant EP/T022159/1), and DiRAC funding from the STFC (www.dirac.ac.uk).

For the purpose of open access, the corresponding author has applied a Creative Commons Attribution (CC BY) licence to any Author Accepted Manuscript version arising from this submission.
\end{acknowledgments}

\section*{Data Availability Statement}

Code for PROFESS-AD can be found in the repository \href{https://github.com/profess-dev/profess-ad}{https://github.com/profess-dev/profess-ad}. The various examples are reproducible by following the tutorial examples in the repository and in the documentation (link in the repository).\\ \\

%\nocite{*}
\bibliography{references}% Produces the bibliography via BibTeX.

\end{document}

\title{Supplementary Material:\\
Automatic Differentiation for Orbital-Free Density Functional Theory}
\author{Chuin Wei Tan, Chris J. Pickard, William C. Witt}

\maketitle

\section{Orbital-Free Density Functional Theory}

\subsection{General Formulation}
As highlighted in the main text, the goal of orbital-free density functional theory software is to compute the ground state energy and density for a given external potential. To ensure proper normalization of the density, it is common to formulate this problem in terms of a Lagrangian
\begin{equation}
\label{lagrangian}
    \mathcal{L}[n] = E[n] - \mu \left[\int n(\mathbf{r}) d^3\mathbf{r} - N_e \right],
\end{equation}
where $n(\mathbf{r})\geq 0$ and $N_e$ is the number of electrons. The ground state density is hence a solution of the Euler-Lagrange equation 
\begin{equation}
\label{euler-lagrange}
    \frac{\delta E}{\delta n}\Bigg\vert_{n_\text{gs}} = \mu,
\end{equation}
where $\mu$ is the chemical potential. While other methods exist for solving the Euler-Lagrange equation,\cite{ofdft_scheme_sle-1,ofdft_scheme_sle-2,ofdft_scheme_sle-3,imoto, oescf} we employ direct minimization of the energy functional.

\subsection{Density Optimization with Unconstrained Variables}

As explained in the main text, we can recast a constrained density optimization problem as a generic unconstrained minimization problem by minimizing the energy functional with respect to an unconstrained variable $\chi(\mathbf{r})$, related to the positive and normalized density by
\begin{equation}
    n[\chi](\mathbf{r}) = \frac{N}{\tilde{N}} \chi^2(\mathbf{r}), ~~\text{where}~~ \tilde{N} = \int \chi^2(\mathbf{r}'')d^3\mathbf{r}''.
\end{equation}
Let us first consider the derivative $\delta n(\mathbf{r})/\delta \chi(\mathbf{r}')$ as defined by
\begin{equation}
    \frac{\delta n(\mathbf{r})}{\delta \chi(\mathbf{r}')} = \lim_{\epsilon \to 0} \frac{d}{d\epsilon} \left\{ n~[\chi(\mathbf{r}) + \epsilon \delta(\mathbf{r}-\mathbf{r}')]  (\mathbf{r}) \right\} 
\end{equation}
Explicit evaluation of the derivative results in
\begin{equation}
    \frac{\delta n(\mathbf{r})}{\delta \chi(\mathbf{r}')} = \frac{N}{\tilde{N}}~ 2 \chi(\mathbf{r}) \delta(\mathbf{r}-\mathbf{r}') - \frac{N}{\tilde{N}^2}~2\chi(\mathbf{r}') \chi^2(\mathbf{r}).
\end{equation}
Hence, we obtain
\begin{equation}
\label{dEdn_to_dEdchi}
\begin{aligned}
    \frac{\delta E}{\delta \chi(\mathbf{r}')} 
    &= \int d^3\mathbf{r} \frac{\delta E}{\delta n(\mathbf{r})} \frac{\delta n(\mathbf{r})}{\delta \chi(\mathbf{r}')}  \\
    &= \frac{N}{\tilde{N}} 2 \chi(\mathbf{r}') \left[ \frac{\delta E}{\delta n(\mathbf{r}')} - \frac{1}{N} \int d^3\mathbf{r} ~\frac{\delta E}{\delta n(\mathbf{r})} n(\mathbf{r}) \right].
\end{aligned}
\end{equation}
This expression allows for transformations between the more commonly known density functional derivatives $\delta E/\delta n$ and the functional derivative required for the unconstrained optimization, $\delta E/\delta \chi$. \\

Additionally, we obtain an analytical form for the chemical potential $\mu = \delta E/\delta n \vert_{n_\text{gs}}$. Noting that $\chi_\text{gs}$ corresponds to $n_\text{gs}$, we can apply the density optimization condition in terms of the unconstrained variable, $\delta E/\delta \chi \vert_{\chi_\text{gs}} = 0$, to Eq.~\ref{dEdn_to_dEdchi} to obtain
\begin{equation}
\label{chemical_potential}
    \mu = \frac{\delta E}{\delta n(\mathbf{r})}\Bigg\vert_{n_\text{gs}} = \frac{1}{N} \int d^3\mathbf{r}' ~\frac{\delta E}{\delta n(\mathbf{r}')}\Bigg\vert_{n_\text{gs}} n(\mathbf{r}') 
\end{equation}
This result holds only when the density optimization is perfectly converged. In practical OFDFT calculations, satisfaction of  Eq.~\ref{chemical_potential} can be used as a measure of convergence. 

\section{Relating Strain Derivatives to Lattice Derivatives}
\label{strain2lattice}
Let the configuration of a continuous system be parameterized by some vector $\mathbf{x}$ with components $x_i$. In the case of crystals, this vector could refer to the lattice vectors or Cartesian ionic coordinates. To describe the deformation from an initial configuration $\mathbf{x}$ to a final configuration $\mathbf{x}'$, we let $\mathbf{x}'$ be parameterized as a function of the original reference configuration such that each element has the dependence $x'_i(\mathbf{x})$. We also introduce the transformation matrix elements $\alpha_{ij}$ and displacement gradient elements $u_{ij}$,
\begin{equation}
    \alpha_{ij} = \frac{\partial x'_i}{\partial x_j} ~~ \text{and} ~~ u_{ij} = \frac{\partial (x'_i-x_i)}{\partial x_j}.
\end{equation}
From these definitions, it is clear that $\alpha_{ij}$ and $u_{ij}$ are related by
\begin{equation}
\label{transformation_matrix}
    \alpha_{ij} = \delta_{ij} + u_{ij}.
\end{equation}
We can hence describe the deformation from $\mathbf{x}$ to $\mathbf{x}'$ in terms of these coefficients,
\begin{equation}
\label{deformation}
    x'_i = \alpha_{ij} x_j = (\delta_{ij} + u_{ij}) x_j.
\end{equation}
The displacement gradient can be decomposed into a symmetric and antisymmetric part. The symmetric part, $\epsilon_{ij}$ is known as Cauchy's infinitesimal strain tensor and measures pure strain while the antisymmetric part, $\omega_{ij}$, measures pure rotation. They are given by
\begin{equation}
\label{strain}
\begin{aligned}
    \epsilon_{ij} 
    &= \frac{1}{2} \left(u_{ij} + u_{ji} \right) = \frac{1}{2} \left(\alpha_{ij} + \alpha_{ji} \right) - \delta_{ij} ~~\text{and}  \\
    \omega_{ij} 
    &= \frac{1}{2}  \left(u_{ij} - u_{ji} \right) = \frac{1}{2} \left(\alpha_{ij} - \alpha_{ji} \right) 
\end{aligned}
\end{equation}
Eq.~\ref{strain} can be manipulated to express $\alpha_{ij}$ in terms of the Cauchy strain and obtain the corresponding partial derivative. 
\begin{equation}
\begin{aligned}
    \alpha_{ij} & = \frac{1}{2} \left(\epsilon_{ij} + \epsilon_{ji} + \omega_{ij} - \omega_{ji} \right) + \delta_{ij} \\
    \frac{\partial \alpha_{mn}}{\partial \epsilon_{ij}} &= \frac{1}{2} \left( \delta_{mi} \delta_{nj} + \delta_{mj} \delta_{ni} \right) 
\end{aligned}
\end{equation}
Defining the lattice vector matrix $h$ such that its columns are the lattice vectors in Cartesian coordinates, Eq.~\ref{deformation} can then be applied on $h$ such that 
\begin{equation}
    h_{k\ell}' = \alpha_{\ell m} h_{m \ell} ~~ \text{and} ~~ \frac{\partial h_{k\ell}}{\alpha_{mn}} = \delta_{km} h_{n\ell}.
\end{equation}
Now consider the derivative of a function of the lattice vectors, $f(\{h_{\mu\nu}\})$, with respect to the infinitesimal strain.
\begin{equation}
\begin{aligned}
    \frac{\partial f(\{h_{\mu\nu}\})}{\partial \epsilon_{ij}} 
    &= \frac{\partial \alpha_{mn}}{\partial \epsilon_{ij}} \frac{\partial h_{k\ell}}{\partial \alpha_{mn}} \frac{\partial f(\{h_{\mu\nu}\})}{\partial h_{k\ell}} \\
    &= \frac{1}{2} \left( \frac{\partial f}{\partial h_{ik}} h_{jk} + \frac{\partial f}{\partial h_{jk}} h_{ik} \right)
\end{aligned}
\end{equation}
The  Cauchy stress tensor and the second-order elastic constants (or Birch coefficients\cite{bm_eos,wallace}) can hence be expressed as
\begin{equation}
\label{stress}
    \sigma_{ij} = \frac{1}{\Omega} \frac{\partial E_\text{gs}}{\partial \epsilon_{ij}}\Bigg\vert_{\epsilon_{ij} = 0} = \frac{1}{2\Omega} \sum_k \left(\frac{\partial E_\text{gs}}{\partial h_{ik}} h_{jk} + \frac{\partial E_\text{gs}}{\partial h_{jk}} h_{ik} \right)
\end{equation}
and
\begin{equation}
\label{elastic_constants}
    C_{ijk\ell} = \frac{\partial \sigma_{ij}}{\partial \epsilon_{k\ell}}\Bigg\vert_{\epsilon_{ij} = 0}  = \frac{1}{2} \sum_m \left( \frac{\partial \sigma_{ij}}{\partial h_{km}} h_{\ell m} + \frac{\partial \sigma_{ij}}{\partial h_{\ell m}} h_{k m} \right).
\end{equation}

\section{Example: Training Parameterized Functionals}

The main text highlighted some AD-based tools for computing additional derivatives of energy functionals. One such quantity is
\begin{equation} \label{lindhard}
    G^{-1}(\eta) = \frac{\pi^2}{k_F} \left( \hat{\mathcal{F}} \left\{ \frac{\delta^2 T_\text{S}}{\delta n(\mathbf{r}) \delta n(\mathbf{r}')} \right\} \Bigg\vert_{n_0}  \right)^{-1} ,
\end{equation}
where $n_0$ is a uniform density, $k_F = (3 \pi^2 n_0)^{1/3}$ is the Fermi wavevector, and $\eta = |\mathbf{q}|/2k_F$ is a dimensionless wavevector, which should be equivalent to the Lindhard linear response function for a free electron gas.\cite{lindhard}\\

We highlight the utility of this feature by considering a mini-problem faced by Imoto et al.\cite{imoto} in the development of their machine-learned semilocal kinetic energy functional. They considered a functional of the PGSL$\beta$\cite{PG1} form, 
\begin{equation}
    T^\text{PGSL$\beta$}_\text{S} = T_\text{vW} + \int_\Omega d^3\mathbf{r}~ F^\text{PGSL$\beta$}_\theta(\mathbf{r}) \tau_\text{TF} (\mathbf{r}) ,
\end{equation}
where the PGSL$\beta$ Pauli enhancement factor is
\begin{equation}
    F^\text{PGSL$\beta$}_\theta(s,q) = e^{-40/27 s^2} + \beta q^2 .
\end{equation}

\begin{figure}[H]
\centering
\includegraphics[scale=0.7]{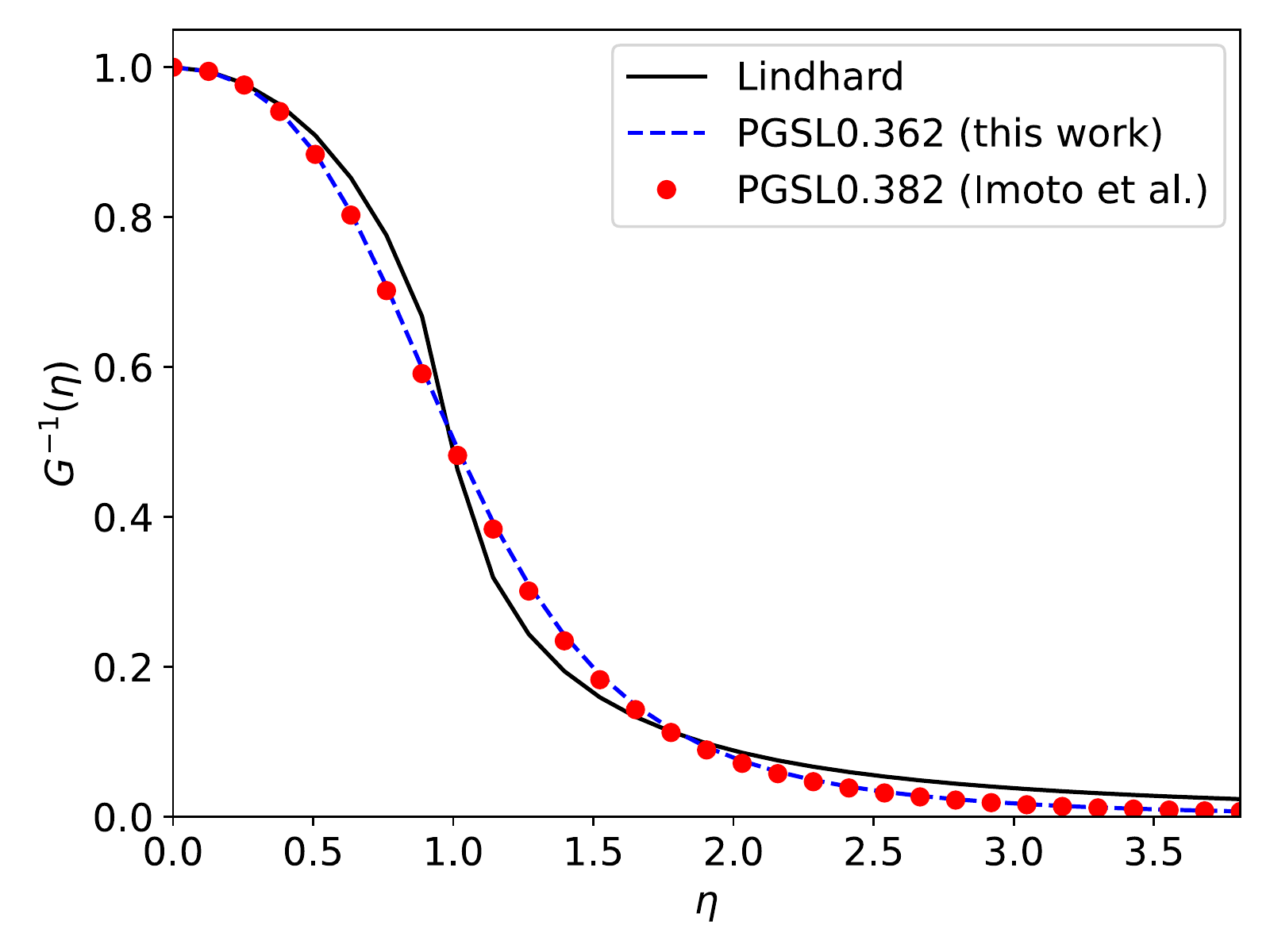}
\caption{Dimensionless linear response functions $G^{-1}(\eta)$ of the PGSL$\beta$ functional\cite{PG1} fitted to match the Lindhard response function. The fit using PROFESS-AD resulted in $\beta=0.362$ while Imoto et al. obtained $\beta=0.382$.\cite{imoto}}
\label{imoto_fit}
\end{figure}

Their goal was to optimize $\beta$ for the functional's linear response to match the Lindhard response function as closely as possible. They performed a least squares fit to minimize the squared deviation between the analytical form of $G_\text{PGSL$\beta$}^{-1}(\eta)$ and $G^{-1}_\text{Lind}(\eta)$ for some range of $\eta$. This gave a result of $\beta = 0.382$. In contrast, PROFESS-AD allows one to perform the same task with minimal human effort. The result of our fit is $\beta=0.362$, which gives very similar $G^{-1}(\eta)$ curves as Imoto et al.'s fit, as shown in Fig. \ref{imoto_fit}. Of course, the PGSL$\beta$ expression used in this example has a linear response function that is relatively easily-derivable. However, the workflow that PROFESS-AD offers can be applied to other functional parameterizations of arbitrary complexity.

%\nocite{*}
\bibliography{references}% Produces the bibliography via BibTeX.